\documentclass[runningheads]{llncs}
\usepackage[T1]{fontenc}
\usepackage{graphicx}
\usepackage{amsfonts}
\usepackage{amsmath}
\usepackage[hidelinks]{hyperref}
\usepackage{booktabs}
\usepackage{subcaption}
\usepackage[disable]{todonotes}
\usepackage{multirow}
\usepackage{graphicx}
\usepackage{fixfoot}
\usepackage{orcidlink}
\usepackage[normalem]{ulem}
\usepackage{float}

\begin{document}
\newcommand{\experimentRepo}{\url{https://github.com/wendli01/missing_link}}

\newcommand{\new}[1]{\textcolor{olive}{#1}}
\renewcommand{\new}[1]{#1}

\definecolor{greenish}{HTML}{82B366}
\definecolor{blueish}{HTML}{6C8EBF}
\definecolor{orangeish}{HTML}{D79B00}
\definecolor{redish}{HTML}{B85450}
\definecolor{bluegray}{HTML}{647687}

\DeclareFixedFootnote{\osf}{\url{https://osf.io/8d2v4/}; March 30, 2025}
\DeclareFixedFootnote{\rgcn}{The large test time difference is due to the optimized RGCN implementation in \texttt{DGL}}

\title{The Missing Link: Joint Legal Citation Prediction using Heterogeneous Graph Enrichment}
\titlerunning{The Missing Link}

\author{Lorenz Wendlinger\orcidlink{0000-0001-9459-6244}\textsuperscript{1} \and
Simon Alexander Nonn\orcidlink{0009-0007-7173-5302}\textsuperscript{1}\and
Abdullah Al Zubaer\orcidlink{0009-0001-0842-7434}\textsuperscript{1}\and
Michael Granitzer\orcidlink{0000-0003-3566-5507}\textsuperscript{1,2}}
\authorrunning{L. Wendlinger et al.}
\institute{\textsuperscript{1} Universit\"at Passau, Passau, Germany \\
\textsuperscript{2} Interdisciplinary Transformation University Austria, Linz, Austria
	\email{lorenz.wendlinger@uni-passau.de}, \email{simon.nonn@uni-passau.de}
        \email{abdullahal.zubaer@uni-passau.de}, \email{michael.granitzer@uni-passau.de}
}

%\author{Anonymous Authors}
%\authorrunning{Anonymous Authors}
%\institute{}
%
\maketitle              % typeset the header of the contribution

\begin{abstract}
Legal systems heavily rely on cross-citations of legal norms as well as previous court decisions.
Practitioners, novices and legal AI systems need access to these relevant data to inform appraisals and judgments.
We propose a Graph-Neural-Network (GNN) link prediction model that can identify Case-Law and Case-Case citations with high proficiency through fusion of semantic and topological information.
We introduce adapted relational graph convolutions operating on an extended and enriched version of the original citation graph that allow the topological integration of semantic meta-information.
This further improves prediction by 3.1 points of average precision and by 8.5 points in data sparsity as well as showing robust performance over time and in challenging fully inductive prediction.
Jointly learning and predicting case and norm citations achieves a large synergistic effect that improves case citation prediction by up to 4.7 points, at almost doubled efficiency.
%We further propose a realistic time-based evaluation methodology that is inherently semi-inductive and operates on valid assumptions.

\end{abstract}
\section{Introduction}

% - link prediction for user assistance

To prepare any legal argument and especially an appraisal, the relevant norms for the specific circumstances of the case must be found, checked, and referenced with concurring or dissenting opinions in the literature and court judgments to support one's (legal) argumentation.
For experienced legal practitioners, who specialize in one area of the law, this context information is available through convention and memorization or by consulting the relevant commentaries.
However, for novice legal practitioners and students, the relevant references can be challenging to research; especially the initial norms for a case can be hard to find.
Some conventions are not obvious without intimate knowledge of the relevant law and legal opinions as well as court decisions in a specific domain.

While legal reference extraction (\cite{palmirani2003automated,shulayeva2017recognizing}) can make this data available, it cannot extrapolate missing references.
The prediction of references exceeds the pure retrieval of similar documents, as it utilizes and extends the reference corpus of a legal text that is inexorably linked to its outcome and logical framework.
It can help legal students and practitioners by suggesting suitable references or indicating incorrect citations for their argumentation.
Furthermore, the so-extracted context can be used for further processing in generative models, making link prediction an essential task for building effective legal tech systems.

Recent advances in LLMs make this problem tractable with generative models, which are more versatile than discriminative models, but can be less robust and tend to hallucinate.
Recent work\,\cite{zhang2025citalawenhancingllmcitations} suggests that they require resource-intensive compute and natural language inference, making them unsuited to resource-constrained and real-time applications.

We solve this task by predicting legal norm and case references for documents with missing references based on semantics and topology \new{to provide a missing link between these data}.
This fusion enables us to include rich meta-information into document representation by modelling them in a joint citation graph.
This meta-data implicitly encodes relevant details about the relevant law domain and makes this important context available even if other data is missing.
Specifically, our contributions are:

\begin{itemize}
    \item We effectively and efficiently predict case and law references in a large citation network composed of real cases and laws by leveraging semantic text-based and topological information
    \item We incorporate semantically relevant categorical meta-information via graph enrichment without the need for separate feature extraction
    \item We modify the original relational graph convolution with a general residual for better representation learning in large sparse graphs
    \item We jointly learn and predict two types of references, drastically improving efficiency and results on smaller graphs
    \item We propose a realistic evaluation method that is temporally cohesive and tailored to a wide range of possible applications
\end{itemize}

% - GNN: general purpose

\section{Related Work}
There is a large catalogue of work on graph learning and related tasks.
While many rule-based approaches to link prediction have been explored in the past\,\cite{arrar2024comprehensive}, they are not easily adaptable to heterogeneous graphs.
We therefore focus on deep learning techniques that can be extended to handle such data, which are commonly known as Graph Neural Networks (\textbf{GNN}).

A link prediction model can be considered a graph auto-encoder as it learns to reconstruct the original graph from representations generated by the encoder.
%In this framework, we can identify the encoder, that generates node representations, and the decoder, that computes the link likelihood between two node embeddings.
A popular choice for the decoder is the dot product for its computational efficiency.
This makes the decoder completely static, while the encoder is learned via back-propagation on the reconstruction task, i.e. binary link classification.

Most GNN encoders are built around the graph convolution (\textbf{GCN}) developed by Kipf et al. in \cite{kipf2016semi}.
They use topological information to compute node-level features in each layer by propagation along graph edges, effectively leveraging the representations of neighbouring nodes.

The Variational Graph Auto Encoder (\textbf{VGAE})\,\cite{kipf2016variationalgraphautoencoders} is an extension that predicts the target distribution and is trained using the Kullback–Leibler divergence in conjunction with the reconstruction loss.

Veličković et al. \cite{velivckovic2017graph} propose the graph attention mechanism (\textbf{GAT}) which computes attention scores along edges.
This includes a learnable notion of node importance into the neighbourhood aggregation.

Schlichtkrull et al. introduce the relational \textbf{RGCN}\,\cite{schlichtkrull2018modeling} for relational graphs that propagates all relations using separate learned weights and then take their normalized sum as the node representation.

However, Wang et al. \cite{wang2021benchmarkinggraphneuralnetworks} show that, in a fair comparison, GCN outperforms these three adaptations in the most common link prediction task.
This partially negates that purported progress of link prediction models and emphasizes the importance of realistic evaluation.

%\subsubsection{Heterogeneous Graphs}

Wang et al. explore Heterogeneous Attention Networks (\textbf{HAN}) for node classification on small heterogeneous graphs in \cite{wang2019heterogeneous}.
They operate on reachable graphs induced on the original network via tuples of relations, so-called metapaths.
This method is not directly adaptable to link prediction, as it removes the original edges, losing valuable information.

%Similar to \cite{wang2021benchmarkinggraphneuralnetworks}, \cite{lv2021we} re-evaluate the results of HAN and related methods and find them to be underwhelming compared to a simpler method.

Dhani et al. explore legal knowledge graphs for link prediction and document similarity in \cite{dhani2021similar}.
Their employ an RGCN with LegalBERT\,\cite{chalkidis2020legal} features on 2\,286 Indian law documents. 
In their link prediction experiments, document embeddings outperform both hand-crafted and bag-of-words node features.

%\cite{xiao2021link}

Palmer et al. \cite{palmer2023re} compare various semantic embedding methods for to Case-Case citation prediction.
They rephrase this task on a paragraph level and show that it is challenging to solve.

%\cite{wendlinger2022reconciliation} use a GCN with bidirectional edges and an asymmetric decoder to predict links in knowledge graphs as well as citation graphs.

%\subsubsection{Decoders}

%Salha et al. \cite{salha2022gravityinspiredgraphautoencodersdirected} compare symmetric and asymmetric decoding methods to their own gravity-inspired decoding in three link prediction tasks and different scenarios. 
%They find that, depending on the reciprocity in the prediction scenario, asymmetric decoding can be vital to making usable predictions.

%\subsection{Link Prediction Evaluation}

%\cite{lichtnwalter2012link} \cite{yang2015evaluating}

\section{Joint Link Prediction with Heterogeneous Graph Enrichment}
\label{sec:abres-gcn}

Our Heterogeneous Graph Enrichment Model (\textbf{HGE}) expands on previous work regarding link prediction on relational data and utilization of meta-features.

We focus on citation networks, a heterogeneous graph structure $\mathcal{G} = (\mathcal{V}, \mathcal{E})$, composed of Nodes $\mathcal{V}$ that represent the topology of documents and the citations $\mathcal{E}$ between them.
Specifically, $\mathcal{V}=\{\mathcal{V}_C, \mathcal{V}_L\}$ contains both \textit{Cases} $\mathcal{V}_C$ and \textit{Laws} $\mathcal{V}_L$, with relations, or edge types, $T_e = \{r_{CC}, r_{CL}\}$.
Cases can reference cases via $r_{CC}: \mathcal{V}_C \to \mathcal{V}_C$ and laws via $r_{CC}: \mathcal{V}_C \to \mathcal{V}_L$. 
An example is given in \autoref{fig:graph_enrichment}.

\subsection{Heterogeneous Graph Convolution}

Like the relational Graph Convolution of \cite{schlichtkrull2018modeling}, we use separate weight matrices, i.e. GCN layers, for each relation type $r\in T_e$.
%This is rooted in the fact that relational GCN cannot be applied to the multi-link prediction problem without fully homogenizing the graph, which results in the unrecoverable loss of node type information.
%Separate link prediction for the two edge types might alleviate this symptom, but, as we will see later, comes with a large performance penality.
For node $i$ in layer $l$, the original RGCN with activation $\sigma$ and normalization constant $c_{i,j}$ computes

\begin{equation}
    h_i^{(l)} = \sigma\bigg( \sum\limits_{r\in T_e} \sum\limits_{J\in \mathcal{N}^r_i} \frac{1}{c_{i,r}} W_r^{(l)}h_j^{(l)} + W_0^{(l)}h_i^{(l-1)}\bigg),
     \label{eq:rgc}
\end{equation}

where $W_r$ is a learnable parameter for each relation and $\mathcal{N}_i^r$ is the neighbourhood of $i$ under $r$.
Unlike \cite{schlichtkrull2018modeling}, we do not model the residual self-loop separately, but instead add a general residual as the self-loop is contained in the enriched graph relation types $ T_e'$ in our model:

\begin{equation}
    h_i^{(l)} = \sigma\bigg(h_i^{(l-1)} + \sum\limits_{r\in T_e'} \sum\limits_{J\in \mathcal{N}^r_i} \frac{1}{c_{i,r}} W_r^{(l)}h_j^{(l)}\bigg),
    \label{eq:hgc}
\end{equation}
\todo{check this}

\begin{figure}
    \centering
    \includegraphics[width=\linewidth]{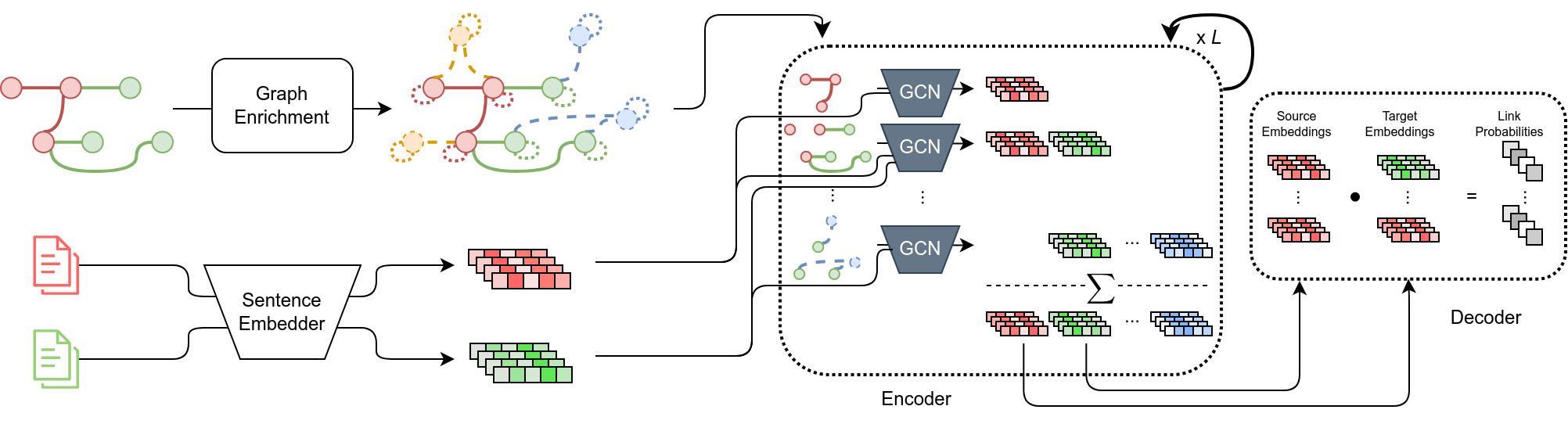}
    \caption{Heterogeneous Graph Enrichment model with \textit{L} layers including and node embedding for \textcolor{redish}{\textit{Cases}} and \textcolor{greenish}{\textit{Laws}}% and graph enrichment with two meta-features (c.f. \autoref{fig:graph_enrichment})
    . 
    All \textcolor{bluegray}{learned} parameters are contained in the encoder. Asymmetric decoding, activation and residuals are omitted for visual clarity.}
    \label{fig:hge}
\end{figure}

Aggregation is still performed for each layer as the sum over all relevant relation types for each node type, c.f. \autoref{fig:hge}.
This can help derive useful features for sparse graphs, such as those of a scale-free nature, by making short feature paths available through residual modelling alongside the long dependencies computed by stacking multiple GNN layers\new{, while alleviating the issue of over-smoothing that plagues deeper GNNs}.
%This allows for more expressive processing, especially in the context of node types with different feature spaces.

\subsection{Joint Modelling}

In contrast to previous work, we investigate link prediction of two distinct edge types.
By jointly learning prediction for both edge types, we can condition the model to produce representations that are informative for both tasks, as Case nodes are used in both relations.
In many graphs, but in our multi-faceted case especially, shortcuts can lead to suboptimal training results \new{that do not generalize well}.
Joint learning achieves a regularizing effect that prevents collapse to the trivial solution.%\footnote{This is strong enough to make inverse frequency loss weighting superfluous}.

\subsection{Graph Enrichment}
\label{subsec:enrichment}

We perform metafeature-based graph enrichment to accurately model the relationship between cases and laws. 
Like in \cite{schlichtkrull2018modeling}, we also add reverse edges and self-loops, though they are processed as regular relations, c.f. \autoref{eq:hgc}.

\paragraph{Exposed Meta-Features}

Descriptive and structural meta-data can provide relevant information about the data itself, as document semantics go beyond just the text.
Especially for a complex legal system, the level of jurisdiction, type of case, or area of law can drastically change how and which references are made.
We want to include these additional semantics in the node representation, though, unlike the initial text-based node features, this meta-information is categorical and therefore not natively compatible with the node feature space.
However, we can include the meta-information about nodes into their representation via the citation network topology by exposing it as discrete nodes for each feature expression.
For a meta-feature $m: \mathcal{V} \to \{0,1\}^M$ with $M$ possible expressions, we modify the relational graph $\mathcal{G} = (\mathcal{V},\mathcal{E})$ such that

\begin{equation}
\begin{split}
    \mathcal{V}' &= \mathcal{V} \cup\{f : \exists u \in \mathcal{V} : m(u) = f\}\\
    \mathcal{E}' &= \mathcal{E} \cup\{(u,m(u))\ \forall u \in \mathcal{V}\}
\end{split}
\end{equation}

\begin{figure}
    \centering
    \includegraphics[width=.8\linewidth]{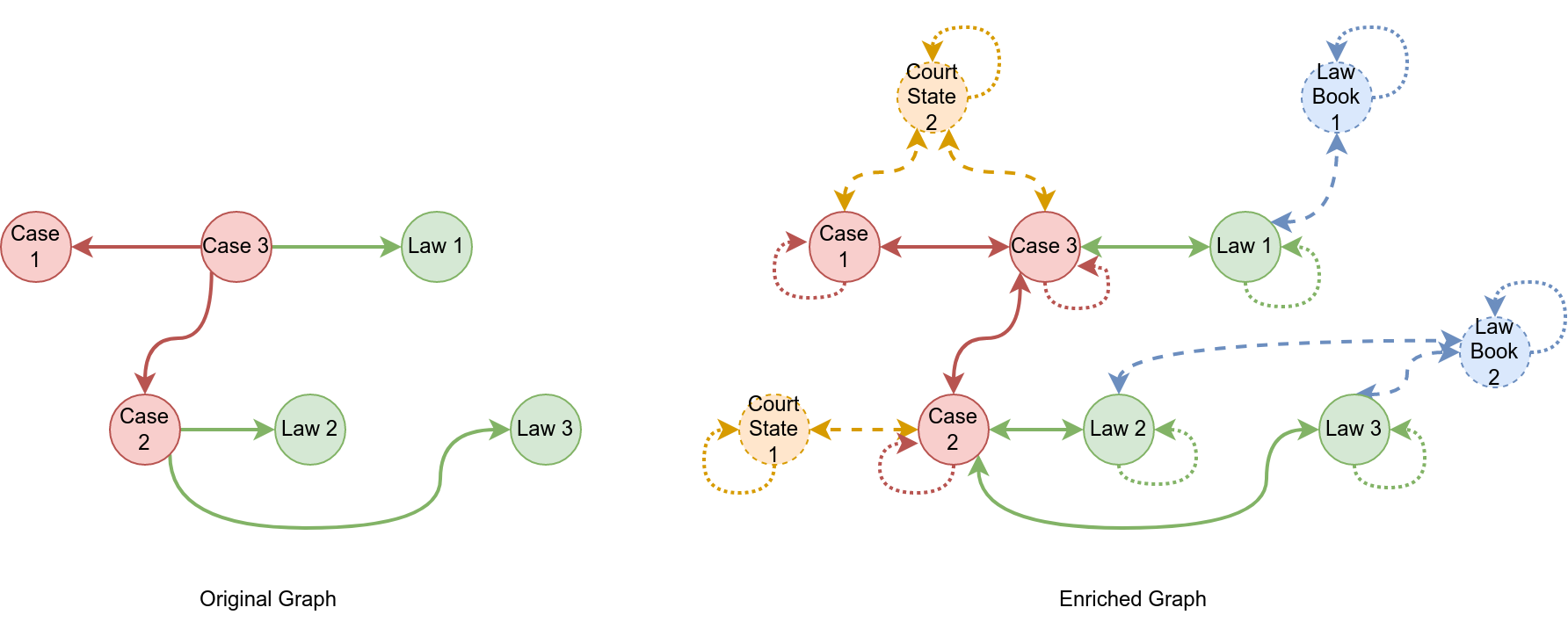}
    \caption{Original and fully enriched graph with dynamic \textcolor{redish}{\textit{Case}} and static \textcolor{greenish}{\textit{Law}} nodes and directed \uline{original edges}. The enriched graph adds \textcolor{blueish}{\textit{Law Book}} and \textcolor{orangeish}{\textit{Case Court State}} nodes with corresponding \dashuline{feature edges} and \dotuline{self-loops} for all nodes.}
    \label{fig:graph_enrichment}
\end{figure}

This does not require additional features for the categories, e.g. courts or law areas.
Moreover, it directly models how nodes that share meta-information are related.
%We experiment with multiple ways of exposing this information (c.f. \ref{subsec:graph_enrichment}), but find that including them as new nodes offers the most versatility.
In this extensional model, a meta-feature is exclusively defined by the nodes that express it, such as an area of law is defined by the laws that govern it via a specific law book, c.f. \autoref{fig:graph_enrichment}.
In contrast to latent aggregation over relevant node representations, we propose dynamically generating it via the same GNN propagation rule.
They are finally aggregated via the same sum rule for heterograph convolution, c.f. \autoref{eq:hgc}.

Intuitively, this serves two purposes: it collects salient representations for exposed features from their respective nodes and propagates them back to the original nodes for contextualizing.
%We experiment with explicitly modelling this feature relationship through metapath processing but find it to be less effective and scalable.
%In the same vein, we try directly using exposed features for contextualization, by normalizing node representations.
%We find that this does not improve results and can lead to unwanted expressions.
We experiment with multiple ways of exposing this information, such as metapath modelling (c.f. \cite{wang2019heterogeneous}) and direct normalization, but we find that including them as discrete nodes offers the most versatility.

%\paragraph{Self Loops}

%Self loops are added as is common for graph convolution to include the node's own representation into its feature computation.
%This extends to all graph enrichment nodes.

%\paragraph{Reverse Edges}

%Reverse edges allow for bidirectional propagation between connected nodes.
%Thereby the representation of a node is informed by both referenced as well as referencing nodes.
%In our graph enrichment scheme, this includes exposed feature edges.

%Treating original and reverse edges as separate link types and processing them separately is necessary to maintain node heterogeneity\footnote{For example, a Case-Law and the reverse Law-Case edge operate on different node types}.
%Unlike the bidirectional propagation rule of \cite{wendlinger2022reconciliation}, we do not learn a recombination rule but aggregate it through the conventional relational accumulation, i.e. sum.

%We show that completely homogenizing the graph to arrive at a natively bidirectional representation does not achieve comparable performance in \ref{subsec:main_study}.

We choose the meta-features \texttt{Case Court Type}, \texttt{Law Code} and \texttt{Case Court State} for graph enrichment.
The \texttt{Case Court Type} is especially informative, as it is composed of both the court level of appeal and jurisdiction.
Together, these meta-features encode information about the most important legal context, i.e. the type of case, area of the law and the local regulations, and make it available explicitly through our graph enrichment.
%They are more expressive than higher-level meta-features, such as court country, but less fragmented than higher cardinality meta-features, such as the court id or court city.

\subsection{Initial Node Features}

Based on the work by Milz et al. \cite{kdir21}, who find strong links between graph and text expression, we propose a fusion of topological and semantic information by integrating textual features into our graph neural network.
We use a BERT-based \cite{devlin-etal-2019-bert} sentence transformer model to generate latent embeddings for the text-based Case and Law nodes.
All legal references are removed via pattern-matching to prevent shortcutting and data leakage.
More specifically, documents are split into context-window size chunks with a small overlap via recursive splitting on white space and then aggregated via their mean.
This ensures maximum context size while maintaining consistency over breakpoints.
We use Jina V2 Embeddings\,\cite{günther2024jinaembeddings28192token} for their large context window of 8192 and availability of German-specific \texttt{jina-v2-base-de} embeddings\footnote{\url{https://huggingface.co/jinaai/jina-embeddings-v2-base-de}}.

We compare the effect of different overlaps and the multilingual Siamese BERT encoder \texttt{all-mpnet-base-v2}\footnote{\url{https://huggingface.co/sentence-transformers/all-mpnet-base-v2}}\cite{reimers-2019-sentence-bert,reimers-2020-multilingual-sentence-bert} in \autoref{subsec:embeddings}.

\subsection{Training Procedure and Parameter Setting}

%Salha et al. \cite{salha2022gravityinspiredgraphautoencodersdirected} compare symmetric and asymmetric decoding methods to their own gravity-inspired decoding in three link prediction tasks and different scenarios. 
%They find that, depending on the reciprocity in the prediction scenario, asymmetric decoding can be vital to making usable predictions. 

To generate link likelihoods from node representations, we use the asymmetric inner product decoder from \cite{yu2014link}.
HGE is trained by optimizing the cross-entropy reconstruction loss via adaptive momentum\,\cite{kingma2014adam} gradient descent %with decoupled weight decay\,\cite{loshchilov2019decoupledweightdecayregularization} 
for 200 epochs with a learning rate of $10^{-4}$.
Random dropout\,\cite{hinton2012improving} is applied to the node representations to prevent co-adaptation and promote robustness
\new{ with a probability of 0.2.
We use 3 layers of size 256 for all models and 2 attention heads for GAT.}
%For the full parameter setting, see \autoref{tab:default_parameters}.

\section{Datasets}
\label{sec:datasets}

We empirically verify our method on legal citation graphs from \cite{Ostendorff_2020} and \cite{kdir21}, and label them according to their \textit{Case} counts as \textbf{OLD36k}\footnote{\url{https://static.openlegaldata.io/dumps/de/refs/}; March 30, 2025} and \textbf{OLD201k}\osf.

Ostendorff et al. \cite{Ostendorff_2020} propose the \textbf{O}pen \textbf{L}egal \textbf{D}ata platform for facilitating open access to legal data.
They collect data for more than 250k German laws and court decisions up to 12/2022 by crawling government websites and trusted services.
They perform reference extraction, named entity recognition and topic extraction for some of them and publish the resulting graph, OLD36k.

\begin{table}
    \centering
    \begin{tabular}{c|c|r|r|c}
    \toprule
        Type & Name & OLD36k&OLD201k& meta-information  \\
        \midrule
        \multirow[c]{4}{*}{Node} & Case & 36\,113 & 201\,823 & type, date\\
        & Law & 10\,304 & 50\,814 &law book code, law book title, section\\
        & \multirow[c]{2}{*}{Court*} & \multirow[c]{2}{*}{-} & \multirow[c]{2}{*}{1\,119} & name, type, slug, city, description\\
        &&&&state, jurisdiction, level of appeal\\\midrule
        \multirow[c]{3}{*}{Edge} & \textbf{C}ase-\textbf{C}ase & 17\,065& 90\,189\dag& -\\
        & \textbf{C}ase-\textbf{L}aw & 424\,862& 971\,625\dag&-\\
        & \textbf{C}ase-\textbf{Co}urt* & -& 201\,823\dag&-\\
        \bottomrule
        
    \end{tabular}
    \caption{Dataset statistics for OLD36k and OLD201k. *: not directly used, only for graph enrichment, c.f. \autoref{subsec:enrichment}, \dag: edge counts differ from those reported \cite{kdir21}, which were generated on preliminary data and have been updated\osf.}
    \label{tab:old36k}
\end{table}

Milz et al. \cite{kdir21} analyse a German citation network based on data from \cite{Ostendorff_2020}.
They use an improved reference extraction and linking method to construct the larger OLD201k citation network.
They further show their data to be scale-free and find node similarity to correlate with text similarity.
They identify reference linking as a pertinent issue for Case-Case citations, as only 16.3\% of extracted references could be added to the network due to missing data.

As OLD36k and OLD201k are extracted from Open Legal Data dumps, they have full text available for all cases and laws.%\footnote{For OLD36k law texts are not included and we scrape them from \url{www.gesetze-im-internet.de} with a hit rate of $\frac{9800}{10304} = 95.1\%$.}.
In addition, both datasets contain meta-information about the cases, laws and courts (directly modelled for OLD201k, indirectly for OLD36k).
\todo{add more information about metadata (distributions?)}

Because both graphs are generated from different data bases with different extraction method, and none of the included identifiers is fully unique, losslessly mapping between the two citation networks is not possible and we treat them as separate samplings from the same domain.

As OLD201k is slightly newer with a cutoff date of 2020-12-10, while OLD36k was created on 2019-02-19, and because it is more complete (c.f. \autoref{tab:old36k}), we use it for the bulk of our experiments.
OLD36k is an almost subgraph of OLD201k with a mismatch of only 28 Cases and elucidates behaviour in scenarios with more limited reference sets or missing data.

\begin{figure}
    \begin{subfigure}[t]{.49\textwidth}
        \centering
        \includegraphics[width=\linewidth]{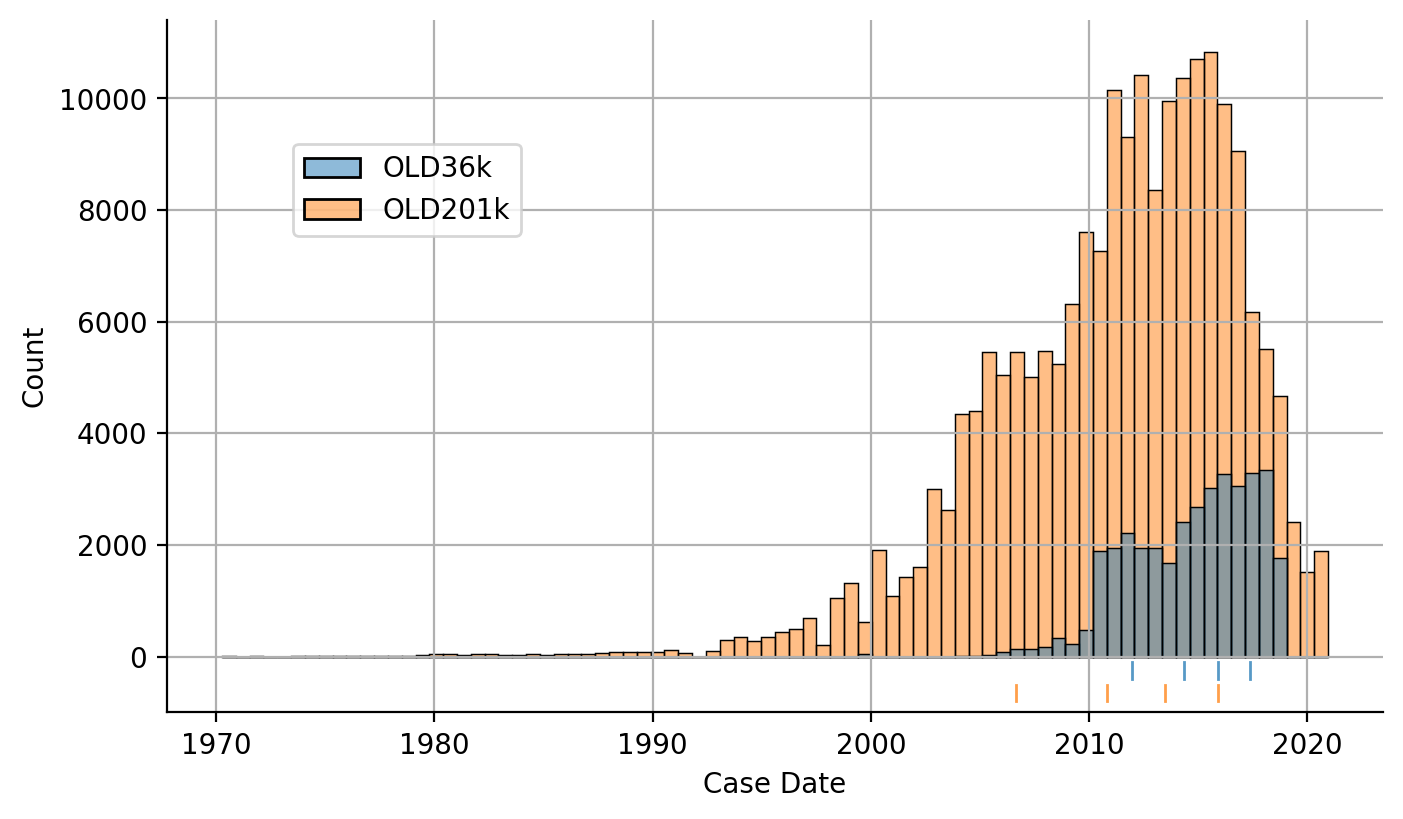}
        \caption{Case date with cutoffs for date splitting indicated for \textcolor{orangeish}{OLD201k} and \textcolor{blueish}{OLD36k}}
        \label{fig:date_dist}
    \end{subfigure}
    \hfill
    \begin{subfigure}[t]{.49\textwidth}
        \centering
        \includegraphics[width=\linewidth]{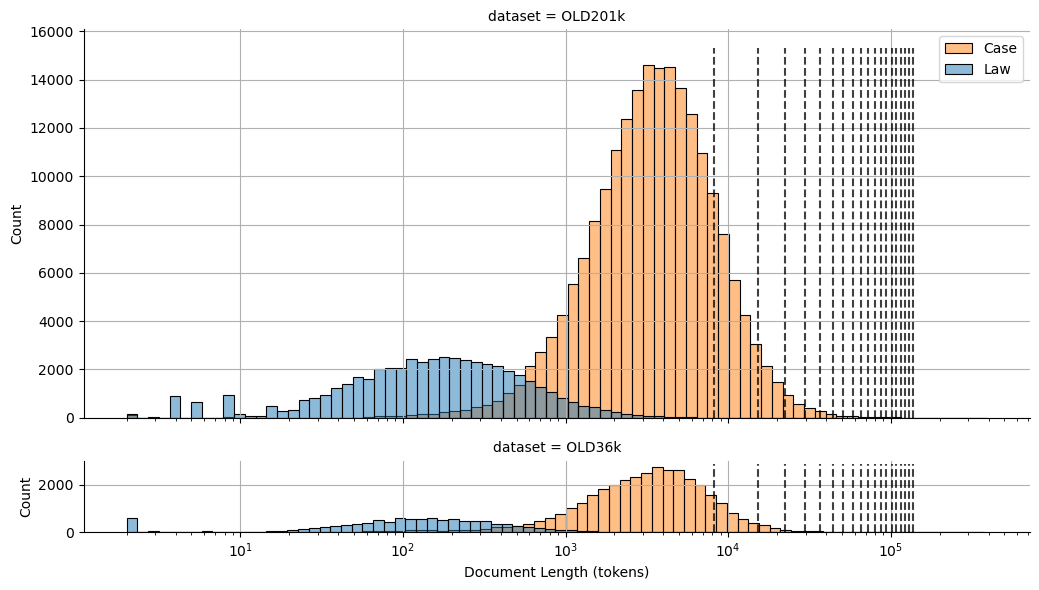}
        \caption{Document Length with embedding context window \dashuline{marked}}
        \label{fig:doc_length}
    \end{subfigure}
    \caption{OLD201k and OLD36k Dataset Distributions.}
    \label{fig:old_dist}
\end{figure}

Their data distributions are similar (c.f. \autoref{fig:old_dist}), though the case library of OLD201k stretches back much further, with some cases from 1970.
OLD201k also contains slightly longer Cases and Laws, with both distributions showing less positive skew.
Consequently, 85\% of OLD201 cases fit within one embedding context window, c.f. \autoref{fig:doc_length}, but 88.9\% of OLD36k cases do.

\section{Experiments}

We first report full link prediction results on OLD201k and OLD36k in \autoref{subsec:main_study} and contrast them with separate prediction in \autoref{subsec:joint_study}.
%Further, we explore the effects of graph enrichment and alternative enrichment methods such as metapath modelling and direct normalization in \autoref{subsec:graph_enrichment}.
An ablation study investigates the effect of various model components in \autoref{subsec:ablation_study}.
The influence of initial node embedding, i.e. text embedding models and chunking parameters, are explored in \autoref{subsec:embeddings}.
Finally, \autoref{subsec:robustness} investigates model robustness and sensitivity in the context of time-based influences and data availability.

We implement all models in \texttt{DGL 2.4.0}\footnote{\url{https://www.dgl.ai/}, March 30, 2025} using \texttt{pytorch 2.6.0}\footnote{\url{https://pytorch.org/}, March 30, 2025} compiled with \texttt{CUDA 12.4} and run experiments on an Nvidia A100 GPU.
All code, experiment setups, final and intermediate results are available at \experimentRepo.

\subsection{Methodology}

Graph splitting is challenging as graphs are a cohesive data structure from which samples cannot be drawn independently.
For link prediction, two main scenarios can be differentiated.
In transductive settings, the node set is the same for training and testing, i.e. all nodes are known at training time, whereas for inductive splitting, predictions are made for new nodes.

\begin{figure}
\centering
    \begin{subfigure}[t]{.34\textwidth}
        \centering
        \includegraphics[width=\linewidth]{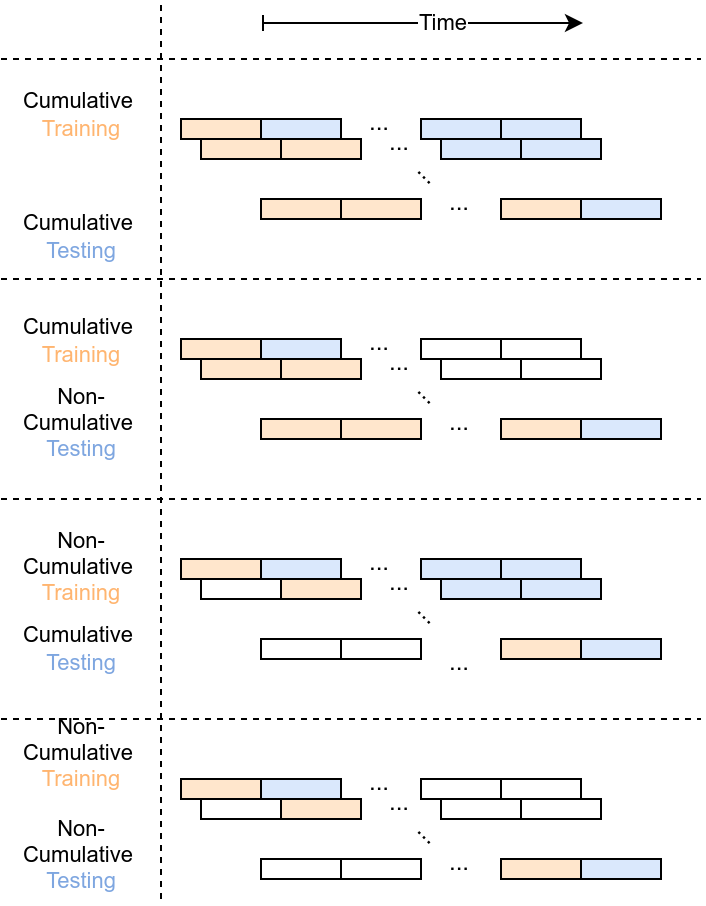}
        \caption{Time-based splitting with (non-)cumulative \textcolor{orangeish}{training} and \textcolor{blueish}{testing}.}
        \label{fig:fig:time-splitting}
    \end{subfigure}
    \hfill
    \begin{subfigure}[t]{.64\textwidth}
        \centering
        \includegraphics[width=\linewidth]{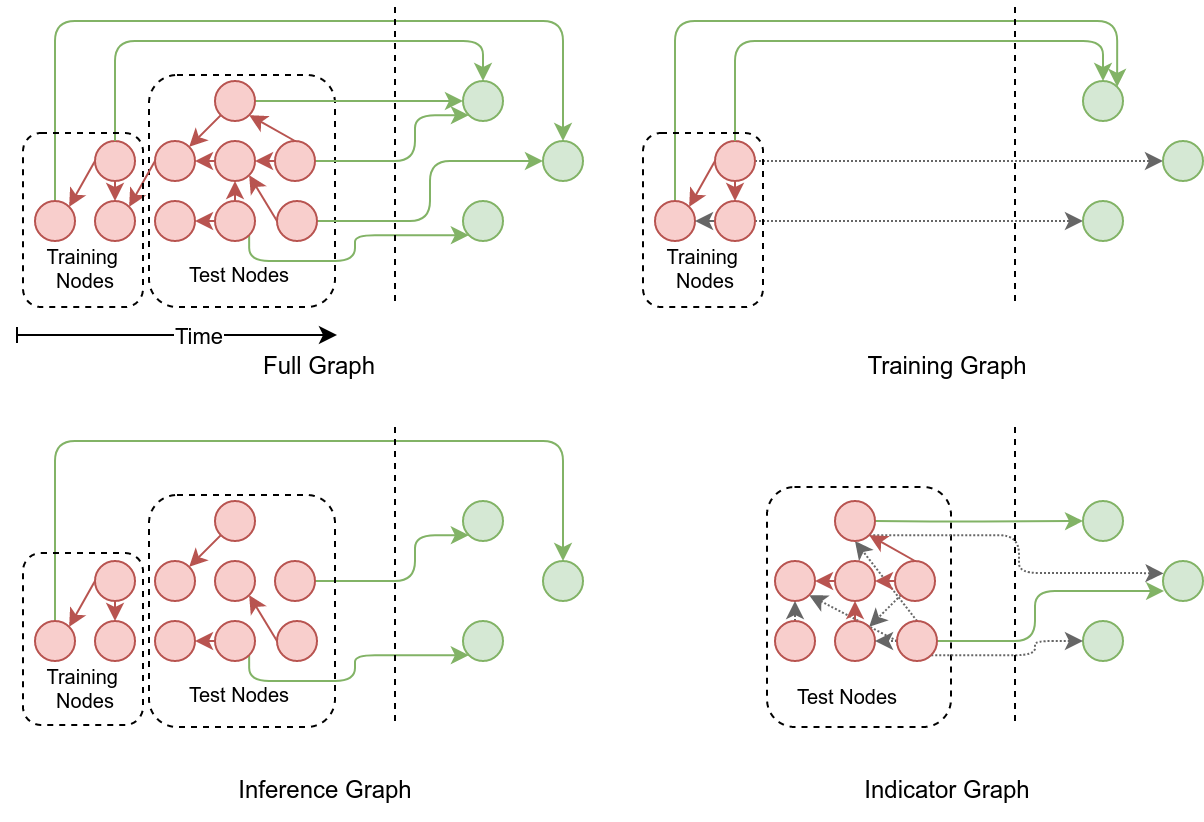}
        \caption{Time-based graph splitting with dynamic \textcolor{redish}{Case} and static \textcolor{greenish}{Law} nodes including \uline{real} edges and training / test \dashuline{negatives}. %The semi-inductive inference graph includes all training data, the indicator graph indicates node pairs to make predictions on.
        }
        \label{fig:graph-splitting}
    \end{subfigure}
    \caption{Time-based splitting modes and graph splitting.}
    \label{fig:splitting}
\end{figure}

Naive random splitting may introduce artifacts by removing nodes that leave their neighbours in an invalid state.
To control this effect, we employ time-based splitting, which ensures a valid training graph for each split, as suggested in \cite{yang2015evaluating}.
This produces a training graph, a semi-inductive inference graph including training data that representations for the test predictions are generated on, and the indicator graph that contains node pairs for prediction, c.f. \autoref{fig:graph-splitting}.
%We choose this cumulative method as suggested in \cite{yang2015evaluating}, as it produces training graphs that were complete at the cutoff time, resulting in only valid assumptions available to the model.

For OLD36k we treat laws as static nodes, while cases are split \new{into equal-sized splits via a cut-off date} based on their associated date, c.f. \autoref{fig:graph-splitting}. 
This automatically results in valid Case-Case edges, as only cases that are already published can be referenced.

By default, we perform cumulative training and testing by training on all old nodes and evaluating on all new nodes, c.f. \autoref{fig:fig:time-splitting}.
This varies the train/test proportion for each split and adds an estimate of data efficiency as well as robustness to concept drift over time.
We evaluate these effects separately in \ref{subsec:robustness}.

By subsampling on both edge types, we can create an incomplete inference graph and a test graph.
This represents a scenario in which complete training data is available, but the current data is missing references.
By default, we create 5 such test splits with a test ratio of 90\% for each time split.
Varying the test ratio allows us to gauge how much information the model can recover without requiring retraining, which we exploit to assess the effects of data sparsity in \ref{subsec:robustness}.
%While this scenario is realistic w.r.t. the graph structure, it still operates on cases that are represented through their full original text and is therefore not causal.
%This means that, even though the actual citation text is removed, the node representation can contain information from passages that describe test citations.
%This data leakage is not representative of the real scenario, in which incomplete case texts are the basis for prediction, and may lead to optimistic results.
%However, identifying problematic text passages, removing them and re-computing node embeddings for each test split is not tractable.
%Moreover, determining how complete removal of a reference changes a legal text is not trivial and might introduce other artifacts.
%We show the effect on a smaller scale in \ref{subsec:robustness}.
For simplicity, we employ balanced negative sampling for both training and testing.
We use uniform per-source sampling, which generates a negative edge $(u, v')\not\in \mathcal{E}$ for each real edge $(u,v)\in \mathcal{E}$.

For each experiment we report the average precision (AP) as well as the area under the Receiver-Operator-Characteristic curve (AUC-ROC).
While AUC-ROC mostly measures the ranking quality of predictions, AP puts a larger emphasis on the positive performance.
These measures are more impactful than metrics that evaluate at a threshold of 0.5, as they assess performance across all possible threshold values.
This is especially relevant in our study, as applications might focus on top-end performance, like recommending the top-n references to a user, or across the whole spectrum, i.e. for retrieval-augmented generation, or at the bottom end for rejecting erroneous user citations.

Scores are calculated separately for the two edge types as well as combined into a micro and macro score.
For imbalanced datasets, macro-averaging ensures that performance on all edge types is adequate, instead of the score being dominated by the majority class, i.e. Case-Law citations, especially as the majority type is usually easier to predict.
This is representative of the application case, where a case citation can be as impactful as a law citation.

\subsection{Link Prediction}
\label{subsec:main_study}

We evaluate the joint link prediction capabilities of our model and previous approaches on  both OLD versions in \autoref{tab:main_study}.
Performance is generally high, owed to the potent fusion of semantic and topological information in our GNN models.
A purely semantic SGD classifier operating only on text embeddings cannot compete.
While for OLD201k homogeneous methods still perform adequately, they cannot recover from the information loss incurred on the homogenized OLD36k.
Their performance is also inconsistent across the two datasets, with VGAE showing promising results on OLD201k but not on OLD36k.

Among the heterogeneous models, our novel HGE out-performs RGCN by a large margin on OLD201k, 2.2 and 3.1 points of average precision and AUC-ROC respectively.
Only HGE can effectively utilize heterogeneous information for improved ranking, while RGCN does not offer AUC-ROC improvements over VGAE.
In the data sparsity of OLD36k, the margin is higher still at 7.2 and 8.5 points respectively.
At the same time, HGE does not add a large performance penalty, with run-time comparable to RGCN\rgcn, as this method also processes each edge type separately, and faster than VGAE, which requires extra standard deviation estimation and loss propagation.
As HGE scales well with increased graph size, it is comparatively more efficient for OLD201k.

\begin{table}
    \centering
    \begin{tabular}{l|ll|ll|ll|ll}
    \toprule
    Data& \multicolumn{4}{c|}{OLD201k} &\multicolumn{4}{c}{OLD36k}\\
    & &AUC- & \multicolumn{2}{c|}{time(s)}& &AUC- & \multicolumn{2}{c}{time(s)}\\
     & AP & ROC &  test & train & AP & ROC &  test & train \\
    \midrule
    SGD & 72.1$_{\pm0.88}$ & 72.6$_{\pm1.11}$ & 4.2$_{\pm1.25}$ & 21.8$_{\pm16.3}$ & 64.4$_{\pm0.22}$ & 64.0$_{\pm0.29}$ & 1.5$_{\pm0.6}$ & 14.9$_{\pm6.6}$ \\
    GAT & 79.6$_{\pm2.75}$ & 81.2$_{\pm2.04}$ & 2.3$_{\pm1.1}$ & 45.0$_{\pm30.3}$ & 73.0$_{\pm0.83}$ & 73.0$_{\pm0.72}$ & 0.7$_{\pm0.9}$ & 19.9$_{\pm2.5}$ \\
    GCN & 80.4$_{\pm1.8}$ & 81.7$_{\pm1.62}$ & 2.3$_{\pm1.1}$ & 44.9$_{\pm30.5}$ & 73.3$_{\pm0.95}$ & 73.3$_{\pm0.94}$ & 0.8$_{\pm01.0}$ & 20.5$_{\pm2.6}$ \\
    VGAE & 82.9$_{\pm1.04}$ & 82.7$_{\pm1.21}$ & 3.3$_{\pm1.1}$ & 111.8$_{\pm69.4}$ & 71.0$_{\pm0.85}$ & 71.7$_{\pm1.0}$ & 1.0$_{\pm1.0}$ & 30.3$_{\pm3.9}$ \\
    RGCN & \uline{85.9}$_{\pm1.46}$ & \uline{82.8}$_{\pm1.99}$ & 0.3$_{\pm0.1}\rgcn$ & 39.2$_{\pm15.9}$ & \uline{80.3}$_{\pm1.4}$ & \uline{77.3}$_{\pm1.27}$ & 0.1$_{\pm0.0}\rgcn$ & 27.1$_{\pm4.9}$ \\
    HGE & \textbf{88.1}$_{\pm1.33}$ & \textbf{85.9}$_{\pm1.55}$ & 3.3$_{\pm0.9}$ & 76.7$_{\pm33.9}$ & \textbf{87.5}$_{\pm1.05}$ & \textbf{85.8}$_{\pm1.34}$ & 1.3$_{\pm1.1}$ & 46.4$_{\pm5.3}$ \\
    \bottomrule
    \end{tabular}
    \caption{Link prediction results in 5 folds with date-based splitting and evaluation on 5 90\% test splits. Results $\pm\sigma$ are reported in \%  with macro-averaging over the edge types, for per-edgetype results see \autoref{tab:joint}. GAT, GCN and VGAE operate on the fully homogeneous bi-directed graph. %, SGD only on node embeddings. 
    %HGE is our fully realized Heterogeneous Graph Enrichment model.
    }
    \label{tab:main_study}
\end{table}

\todo{add topology only results for important models (GCN, RGCN, HGE)?}
\todo{add more meaningful semantic only results (MLP instead of SGD)?}

The promising AP and AUC-ROC scores of HGE on both OLD graphs show that our relational modelling and enrichment is effective and make it suitable for link prediction across a variety of possible thresholds and application scenarios.

\subsection{Joint Prediction}
\label{subsec:joint_study}

As we predict jointly by default, we compare the efficacy and effectiveness to separate prediction here.
The input data is the same for both scenarios, only the prediction target changes - separate models therefore have access to all edges for propagation.
For joint prediction, \textit{CC} and \textit{CL} are learned and predicted by the same model, while separate models are just that.

\begin{table}
    \centering
    \begin{tabular}{ll|ll|ll|ll}
    \toprule
     dataset& training  & \multicolumn{2}{c}{Average Precision} & \multicolumn{2}{|c}{ROC-AUC}& \multicolumn{2}{|c}{time (s)}\\
     &&  CC & CL &  CC & CL  &train&test \\
    \midrule
    \multirow[c]{2}{*}{OLD201k} & joint & \textbf{78.2}$_{\pm2.63}$ & 98.0$_{\pm0.1}$ & \textbf{73.8}$_{\pm3.03}$ & 98.0$_{\pm0.12}$ & \textbf{76.7}$_{\pm34}$ & \textbf{3.3}$_{\pm0.9}$ \\
                             & separate & 77.6$_{\pm3.12}$ & 98.0$_{\pm0.1}$ & 73.4$_{\pm3.18}$ & 98.0$_{\pm0.1}$ & 137.4$_{\pm63}$ & 7.1$_{\pm1.1}$ \\
\cline{1-8}
    \multirow[c]{2}{*}{OLD36k} & joint & \textbf{79.8}$_{\pm2.03}$ & 95.2$_{\pm0.18}$ & \textbf{76.8}$_{\pm2.56}$ & 94.7$_{\pm0.25}$ & \textbf{46.4}$_{\pm5.3}$ & \textbf{1.3}$_{\pm1.1}$ \\
                                & separate & 75.1$_{\pm1.2}$ & 95.1$_{\pm0.22}$ & 69.4$_{\pm1.36}$ & 94.7$_{\pm0.26}$ & 60.3$_{\pm3.4}$ & 1.6$_{\pm1.2}$ \\
    \bottomrule
    \end{tabular}
    \caption{Edge type scores for HGE in joint vs. separate prediction.}
    \label{tab:joint}
\end{table}

Regardless of dataset, \textit{CC} edges are more challenging to predict than \textit{CL}, though for OLD201k the difference is more pronounced.
While this may have inherent reasons, such as higher heterogeneity and time-sensitivity of case citations, it can also be explained by the significant linking issues for CC edges reported in \cite{kdir21}, c.f. \autoref{sec:datasets}.

We find that joint prediction offers regularization that prevents short-cutting and substantially improves results for the minority edge type, \textit{CC}.
This is especially pronounced for OLD36k, where fewer data is available and joint learning increases AP and ROC-AUC by 4.7 and 7.4 points, c.f. \autoref{tab:joint}.
For OLD201k the improvement is still measurable at 0.6 and 0.4 points.
More importantly, due to the joint nature, training is sped up by a factor of 1.79x and 2.15x respectively, with testing scaling similarly.

\subsection{Ablation Studies}
\label{subsec:ablation_study}

We also evaluate the precise impact of different HGE components on link prediction on
OLD201k in \autoref{tab:ablation_study}, while holding all other hyper-parameters constant.

We find that type information has the greatest impact on average precision, though HGE operating on the homogeneous graph still out-performs all other approaches, including the RGCN \new{(\autoref{tab:main_study}), as it can recover a portion of the type information through the semantic node features}.
The effect of reverse edges is comparatively small, and at 0.5 points macro AP, comparable to the 0.7 points macro ROC-AUC difference of omitting residuals (c.f. \autoref{tab:ablation_study}).
Not exposing meta-features lowers AP and ROC-AUC scores by 0.7 and 1 point(s).

\begin{table}
    \centering
    \begin{tabular}{l|ll|ll|ll}
    \toprule
    & \multicolumn{2}{c}{Average Precision} & \multicolumn{2}{|c}{ROC-AUC}&\multicolumn{2}{|c}{time (s)}\\
     & micro & macro & micro & macro&test&train \\

    \midrule
    HGE & 97.1$_{\pm0.12}$ & 88.1$_{\pm1.32}$ & 97.0$_{\pm0.12}$ & 85.9$_{\pm1.54}$ & 3.2$_{\pm0.9}$ & 74.7$_{\pm31.4}$ \\
    w/o reverse edges & 96.1$_{\pm0.27}$ & 87.6$_{\pm1.69}$ & 95.8$_{\pm0.28}$ & 85.9$_{\pm2.31}$ & 2.8$_{\pm0.9}$ & 49.8$_{\pm20.4}$ \\
    w/o skip connections & 96.9$_{\pm0.1}$ & 87.6$_{\pm1.44}$ & 96.8$_{\pm0.13}$ & 85.2$_{\pm1.84}$ & 3.2$_{\pm0.9}$ & 75.4$_{\pm32.5}$ \\
    w/o exposed features & 96.9$_{\pm0.09}$ & 87.4$_{\pm1.5}$ & 96.8$_{\pm0.15}$ & 84.9$_{\pm1.8}$ & 2.9$_{\pm1.8}$ & 51.5$_{\pm24.6}$ \\
    homogeneous & 97.0$_{\pm0.07}$ & 85.7$_{\pm1.43}$ & 97.0$_{\pm0.06}$ & 85.9$_{\pm0.96}$ & 2.6$_{\pm0.9}$ & 34.5$_{\pm10.6}$ \\
    \bottomrule
    \end{tabular}
    \caption{HGE ablation study for OLD201k link prediction in 5 folds with date-based splitting and evaluation on 5 90\% test splits.}
    \label{tab:ablation_study}
\end{table}

We can deduce that reverse edges and type information mostly benefit the identification of positives, while skip connections and exposed meta-features also help with finding representations that can encode negatives.

With everything else equal, as per the HGE results without exposed features, our adapted heterograph convolution still outperforms the original RGCN by 4.6 and 2.9 points of AP and ROC-AUC respectively (c.f. Tables \ref{tab:ablation_study} and \ref{tab:main_study}).

%\subsection{Graph Enrichment}
%\label{subsec:graph_enrichment}

\subsection{Node Embeddings}
\label{subsec:embeddings}

We investigate the impact of the initial node embedding method, as it generates the representation space that nodes are mapped into for all our experiments. 

\begin{table}
    \centering
    \begin{tabular}{l|r|r|ll|ll}
    \toprule
    &context & overlap & \multicolumn{2}{c}{Average Precision} & \multicolumn{2}{|c}{ROC-AUC}\\
     &  &  & micro & macro & micro & macro \\
    \midrule
    \texttt{all-mpnet-base-v2}\,\cite{reimers-2020-multilingual-sentence-bert} & 384 & 16 & 96.6$_{\pm0.2}$ & 86.8$_{\pm1.47}$ & 96.5$_{\pm0.22}$ & 84.3$_{\pm1.96}$ \\\hline
    \multirow[c]{3}{*}{\texttt{jina-V2-base-de}\,\cite{günther2024jinaembeddings28192token}}&4096 & 512 & 97.1$_{\pm0.08}$ & 88.0$_{\pm1.4}$ & 97.0$_{\pm0.09}$ & 85.7$_{\pm1.67}$ \\
    &2048 & 256 & 97.0$_{\pm0.12}$ & 88.0$_{\pm1.44}$ & 96.9$_{\pm0.13}$ & 85.7$_{\pm1.79}$ \\
    &8192 & 1024 & 97.1$_{\pm0.09}$ & 88.1$_{\pm1.34}$ & 97.0$_{\pm0.12}$ & 85.9$_{\pm1.56}$ \\
    \bottomrule
    \end{tabular}
    \caption{Node Embedding study on OLD201k using the full HGE.}
    \label{tab:my_label}
\end{table}

Overall, we observe the impact of the semantic node embedding method to be small, with chunking parameters showing no significant effect (c.f. \autoref{tab:my_label}).
The over-smoothing usually encountered when aggregating over many embeddings is not detrimental.
Even with the substantially smaller embedding and context window of \texttt{all-mpnet-base-v2}, most relevant information can be recovered by HGE.
While we did experiment fine-tuning with Attention over Sentence Embeddings\,\cite{abdaoui2023attention}, it did not improve results and could not easily be scaled to our data.
This confirms our decision of using latent initial node embeddings.
\todo{Add AoSE results and run times?}

\subsection{Robustness and Sensitivity}
\label{subsec:robustness}

In the time-based evaluation scenario we have chosen, multiple aspects can be varied to gauge robustness and sensitivity to temporal and data changes.

\begin{figure}[ht]
    \begin{subfigure}[t]{.49\textwidth}
        \centering
        \includegraphics[width=\linewidth]{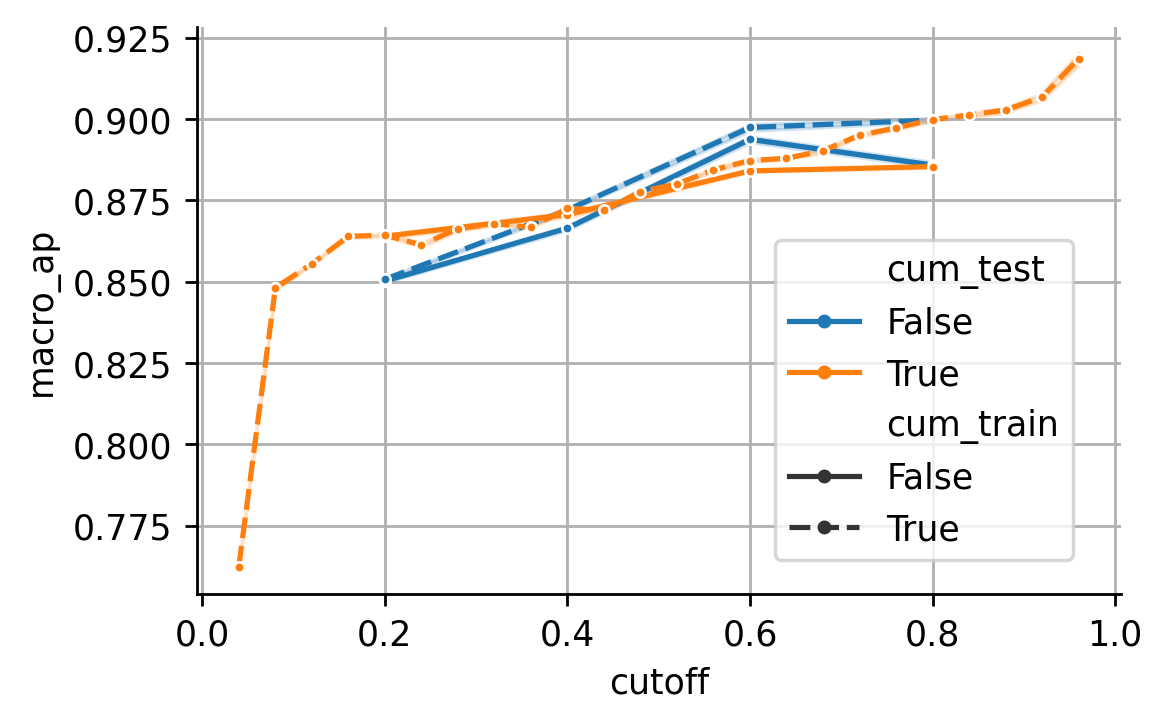}
        \caption{Time and Quantity via date splitting from 1970 to 2022 with (non) cumulative training and testing.}
        \label{fig:data_quantity}
    \end{subfigure}
    \hfill
    \begin{subfigure}[t]{.49\textwidth}
        \centering
        \includegraphics[width=\linewidth]{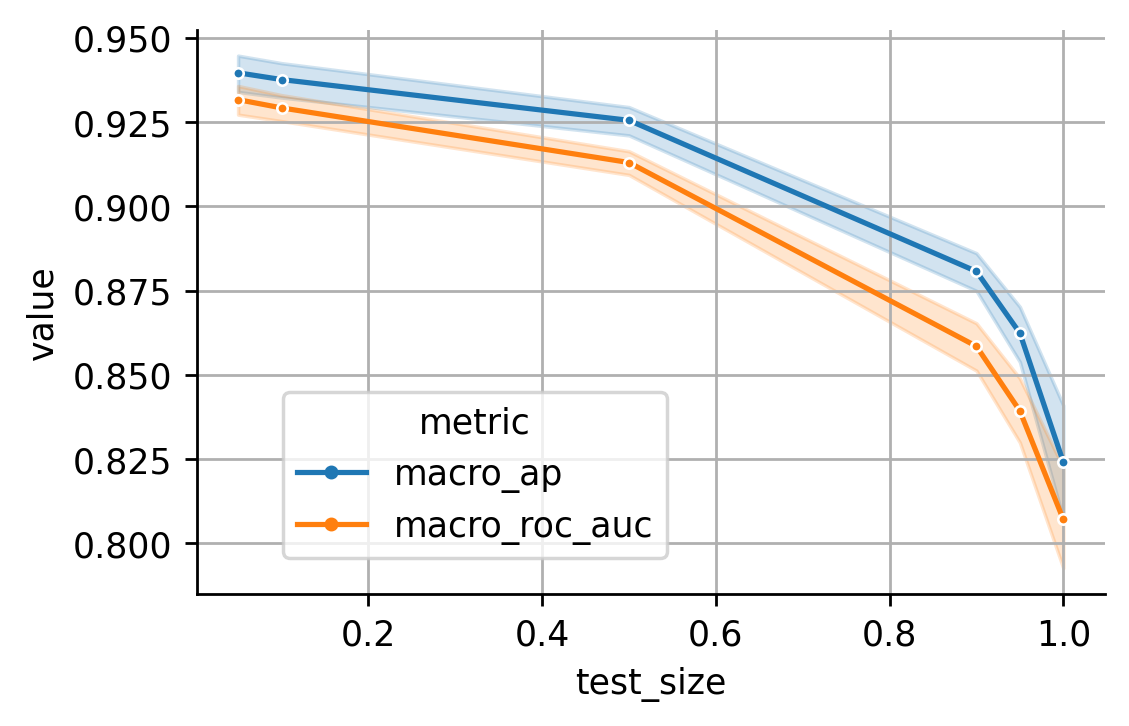}
        \caption{Data Sparsity via test graph edge ratio. At 1.0 prediction is fully inductive.}
        \label{fig:data_sparsity}
    \end{subfigure}
    \caption{Robustness of HGE on OLD201k over time and varying data availability.}
    \label{fig:robustness}
\end{figure}

While performance does improve over time, c.f. \autoref{fig:data_quantity}, this is mostly due to the low data quality of the earliest cases that include only very few extracted citations.
Conversely, this low-quality data does not poison the cumulative training set, as non-cumulative training performance is worse, and HGE does not experience degradation due to drifts.
Non-cumulative testing confirms that there is some temporal dependency in the data, though it is not well-behaved.
\todo{rephrase?}

We estimate the model to be somewhat sensitive to data sparsity in the availability of inference edges, c.f. \autoref{fig:data_sparsity}.
With lower test sizes than the default of 0.9, and correspondingly more inference edges available, model performance is improved.
In the fully inductive setting without any inference links (i.e. test size at 100\%), HGE can still recover information with scores above 80\%, by leveraging the topology induced through our meta-feature graph enrichment.

\section{Conclusion \& Future Work}

We solve the task of legal reference prediction on two large real-world citation networks with a semantic-topology fusion GNN approach that outperforms semantic methods by a large margin.
We further present a novel method for heterogeneous citation prediction that can effectively and efficiently exploit meta-information and synergies in joint learning.
It is robust over a large temporal range and handles data sparsity well.
While it operates on full text features, these are latent and the model is discriminative, which results in promising scalability.

We believe that this can be an effective tool for the writing and assessing of legal documents for both legal practitioners and novices by lowering the barrier of entry for finding norm and case citations for a relevant case.
Moreover, it can help identify even obscure relevant precedent, which is an important step in building legal tech systems.

%\section{Future Work}

%We did not fully re-implement Simple-HGN, as we did not find l2-normalization to be helpful.
%Our experiments did not extend to a homogeneous subgraph with the proposed edge-type embeddings, though \cite{lv2021we} find their benefit to be small for link prediction.
%Again, here could lie a combined solution with full Simple-HGN for some and GAT for other edge types.
Though our system is effective, robust and scalable, there are limitations to its applicability, such as the availability of new legal data.
Integrating new knowledge requires embedding cases, extracting references and re-training the model, which we estimate at a runtime of less than one hour (c.f. \autoref{tab:joint}).
Predicting the references of one case requires encoding it and querying all possible targets, i.e. computing the dot-product with all other nodes, which we measure to take less than 5s for OLD201k, though with an optimized implementation, or reference set partitioning, this could be reduced even further.

%As our model is discriminative, it requires evaluation of all relevant node pairs, which usually extends to the whole reference pool.
%This is in contrast to generative models, which are designed to decompose this lookup by auto-regressively predicting tokens instead of discrete references.
%It remains to be seen whether this can be alleviated by exploiting the time stability of our model and defining heuristics to partition the reference pool.

Both considered citation graphs have missing links due to imperfect reference extraction and linking, which, we believe, could be improved by applying new Natural Language Inference methods and limiting the reference set to disjoint graph components.
Identifying them is feasible along jurisdiction and geographic lines with the help of legal experts and would also improve link prediction effectiveness and efficiency by limiting the considered reference pool.

Our reference prediction is limited to the document level throughout, while for legal practitioners, it might be valuable to know which part of their opinion should reference which legal paragraph or case detail, as this can drastically change the outcome of a legal appraisal.
Re-framing this along the lines of \cite{palmer2023re} would require either a separate fine-grained prediction model or modelling as paragraph nodes, with corresponding sentence-level attention, and might only be tractable with a limited reference set, as mentioned above.

As a simplification, we have treated the law as a static document set, though in practice it is ever-changing.
Collecting and integrating previous versions into our reference set could improve performance on and transfer from historic cases.

\subsubsection*{Acknowledgements}
The paper has been partially funded by COMET K1- Competence Center for Integrated Software and AI Systems (INTEGRATE) within the Austrian COMET Program and by the German Federal Ministry of Education and Research (BMBF) within the project DeepWrite (Grant. No. 16DHBKI059).

\bibliographystyle{splncs04}
\bibliography{main}

\iffalse
\section*{Appendix}

\begin{table}[H]
\renewcommand\thetable{A.1}
	\centering
	\begin{tabular}{l|c|c}
        \toprule
		Parameter Name  				& Default Value& GAT  \\
        \midrule
		GCN Layer Sizes					&\multicolumn{2}{c}{(256, 256, 256)}\\\hline
            Attention Heads                 &-&2\\\hline
		Epochs							&\multicolumn{2}{c}{200}	\\\hline
		Dropout	Probability				&\multicolumn{2}{c}{0.2}\\\hline
		Learning Rate			        &\multicolumn{2}{c}{0.01}	\\
        \bottomrule
	\end{tabular}
	\caption{Default hyper-parameters and model-specific settings.}
	\label{tab:default_parameters}
\end{table}
\fi

\end{document}